# The effects of having lists of synonymies on the performance of Afaan Oromo Text Retrieval system

*Isayas W. Kelbessa*

College of Computing and Informatics, University of wolkite

Email: isa4wak3stu0@gmail.com

## Abstract

Obtaining relevant information from a collection of informational resources in Afaan Oromo is very important for Afaan Oromo speakers, developing a system that help users of Afaan Oromo is mandatory. That is why this study is envisioned to make possible retrieval of Afaan Oromo text documents by applying techniques of modern information retrieval system. In the developed Afaan Oromo prototype, Probabilistic approach was used as an information retrieval models and precision and recall measurement were used as the performance measurement or evaluation technique. Apache Solr was also used as an environmental programming language to achieve the evaluation goal. Afaan Oromo text retrieval is evaluated using 158 documents and 13 arbitrarily selected queries that can determine the effectiveness of retrieval using the precision-recall. The average result obtained by our evaluation before the addition of synonymy was 72.91% precision and 86.8% recall respectively. After the addition of synonymy, the value was changed to 71.39% average precision and 90.5% average recall. The F-measure for the evaluation before synonymy addition was 79.25% and after addition changed to 79.82%. The addition of synonymy improves the system performance by 0.57%. The study therefore, experimentally proves that the addition of the thesaurus system can improve the system performance. Spellchecking, pagination, hit highlighting and autosuggestion is also possible in the developed prototype for Afaan Oromo.

***Key words:*** *Afaan Oromo, Apache solr, effect, Information Retrieval, Precision, Probabilistic approach, Recall, Synonymy,*

# 1. Introduction

A Thesaurus is a reference work that enlists words grouped together according to similarity of meaning (containing synonyms and sometimes antonyms), in contrast to a dictionary, which provides definitions for words, and generally lists them in alphabetical order. A thesaurus is a set of terms that are semantically related [1].It helps in improving the quality of retrieval by guiding indexers and searchers about which terms to use. It is a structure which supports the automatic indexing and retrieval [2]. According to [3], Information retrieval is defined as the process of searching a collection of documents, using the term document in its widest sense, in order to identify those documents which deal with a specific subject. It is the recall and the precision which attempt to measure the effectiveness. The information retrieval has changed dramatically in recent years, with the immense increase in availability of searchable full text and the increasing availability of powerful engines for searching the text. Today it is beginning to seem as if all information is available in full text. However, there are many problems (ambiguity, synonyms, etc.) that indicate otherwise. Full text searching will always be valuable for browsing in any size of file but in large files, controlled language access searching will always support efficient retrieval system. If there are some lists of synonymy for any collection of documents when we want to search some documents in Afaan Oromo text retrieval systems, if the exact word is not found in the collection the synonymy of that word may be exist and the Information Retrieval system can return the documents for the user. We have done research on this topic because of two reasons:

1. In our daily life, sometime it's difficult to memorize a word; hence synonymy plays an important role in such conditions.
2. Sometimes, certain words don't fit in a sentence or make the sentence strange and their might be a grammatical error so in that case we can use synonymy to make a good sentence structure or a logical and syntactical sense.

Therefore, we can conclude that thesauri (specifically the paper focused on synonymy) in Information Retrieval Systems and indexing are required in facilitating information retrieval for any language [3].

# 2. Problem statement

Afaan Oromo belongs to the Cushitic branch of the Afro-Asiatic languages together with Afar, Somali and Sidama from more than 80 languages of Ethiopia. Afaan Oromo is one of the languages with large number of speakers under Cushitic family [4]. If Arabic is counted as African languages, Afaan Oromo is the 4th largest language but the 3rd next to Kiswahili and Hausa if Arabic is not an African language [4]. As there are more users of this language, it needs to be able to support technology; because, Afaan Oromo language is used as the official language for oromia regional state of Ethiopia and also academic language for the oromia state.

There is an attempt in Ethiopia, to develop information retrieval system for the official language of Ethiopia which is Amharic. There is also another attempt for Afaan Oromo but most of the attempt is on dictionary based approach instead of corpus based approach. The main important problems that gave rise to conduct this study is that many Oromo people in Ethiopia and also in others country such as: Kenya, Djibouti Somalia, Egypt and South Africa use Afaan Oromo language. Users who look for Afaan Oromo document with Afaan Oromo query may not find suitable environment to find information for their need. The less knowledgeable ones still cannot understand the English language fully. So, why are those people who do not understand English not helped? Why the Oromo people cannot make their language part of the technology's language? Generally, the statements of the problem for the study are the following shortly:

- Many Afaan Oromo speakers nowadays search for their information needs on Google; which is very broad to get what they need. So, the best way is visiting specific sites to get real information they need.
- The rich literacy works in Afaan Oromo language are not accessible in digital form
- To help users of Afaan Oromo to find information they need without any difficulties
- To satisfy users who look for Afaan Oromo document with Afaan Oromo query.

**Research question**

1) What are the effects of synonymy on Afaan Oromo Text Retrieval system?

# 3. Objective of the study

**General objective**

The general objective of this study is to know the effects of having the lists of synonymy in Afaan Oromo Text Retrieval

**Specific objectives**

- To know how lists of synonymy of Afaan Oromo corpus have effect on the performance of the system
- To evaluate the performance of the proposed prototype by using the recall and precision measures and recommend future research direction.

# 4. Scope and limitation of the study

**Scope of the study**

The scope of this study is to know the effects of having the lists of synonymy in Afaan Oromo Text Retrieval system in Afaan Oromo text corpus by using probabilistic approach of IR model. Any other data types, such as image, video, and audio are out of focus of the study.

**Limitation of the study**

In the study, limited corpus was used for evaluating the performance of the system developed as a result of time factor.

## Significance of the study

The speakers of the specific language can benefit from the development of technology in many ways when the language is able to grow with technology.

The study has advantage for:

a) **For the speakers of Afaan Oromo**: - This research helps the speakers of the language as they use their own language to search the information they need, i.e. the query is in Afaan Oromo.

b) **For the Development of the language**: The study also helps the language as it becomes the language of technology which is very important for the development of this language. Identify the best information retrieval model for the language, measures the performance of the recall and precision of the language, the way to preprocess in Afaan Oromo and identifies their difficulties for the next researcher.

# 5. Literature review

According to [3] , Information systems may include the word ¨thesaurus¨ on a navigation bar or as a hypertext button, but the explanation of how this feature can assist with the selection of search terms may be hidden. Electronic thesaurus versions have strengthened its role as a search aid. Many operational systems accessible via the internet have incorporated thesauri in their interface as a part of their browsing and searching facilities.

## Related Works

Many researches have been done on text retrieval in the world and in African countries but few in Ethiopia. However, some work done is found on the specific area of this study from the reviewed literatures, specifically on the bilingual information retrieval (Afaan Oromo-English cross lingual information retrieval) [5]. But, one scholar also made attempt to develop Information Retrieval system that works for Afaan Oromo by using vector space model of Information Retrieval model and python as developmental tool with the corpus size of 100 text documents written in Afaan Oromo language [6]. But there is no information that shows the effect of the lists of synonymy in the system performance for Afaan Oromo.

# 6. Methodology

## Sources of data

Related documents, secondary data, which are specifically related to the study's objectives, have been reviewed from different sources (books, journals, internet, and other thesis) etc. to understand the effects of having lists of synonymy and polysemy in Afaan Oromo text retrieval using probabilistic approach and measure the performance before and after adding the lists.

## Corpus

The corpora have been prepared from different websites such as: the official website of oromia national regional state, website of voice of America Afaan Oromo language, oromia media network site, TV oromia, international bible society official website, oromia radio and television organization (ORTO) and any other Internet based sources. Resources from holy bible books, news articles and the Internet are used as data set. The corpus size is 158 documents. This data set covers different aspects of life like: educational, religious, political, socio-economic, culture, sports and so on. Information retrieval needs standard corpus for testing and making experimentation.

## Developmental tools

In this study a single personal computer processor of intel core i5 with 2.4GHz speed, Microsoft windows 10 professional operating system, 500GB Hard Disk capacity and 4GB of RAM is used. For the development of the prototype, we have utilized the following tools.

**JDK 1.8.0.91:** Since all components of our apache solr is developed using java programming language, JDK (java development kit) version 1.8.0.91 is utilized and installed on windows environment for implementing the various components of the solr.

**NetBeans 8.0.1:** we have used netbeans to implement BM25 similarities in apache solr. It used for writing, running and compiling any java codes that we need to connect to lucene.

**Apache solr (solr-6.0.0):** Apache Solr was used as windows environment and programming language for the purpose of this research. Apache solr support both *query effectiveness* and *query efficiency* [5] which motivate the researcher of this study to use this tools.

## Experimentation and Discussion

To test this study experimentally, the performance of the IR system using corpus which is called in our case *Afaan-Oromo-core* is used. We have collected data from different websites that contain Afaan Oromo words and statements. This corpus also contains different formats of data which seems like the following.

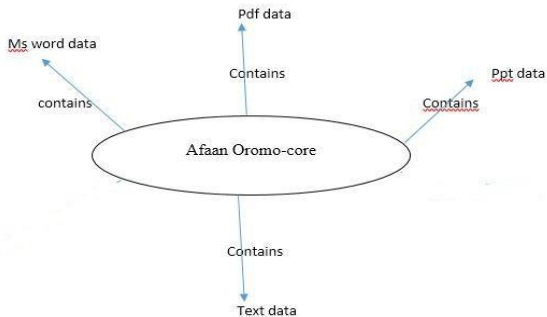

**Figure 1:** Types of data used for Afaan Oromo Corpus

The data which is in (Microsoft word, power point, portable digital format, extensible markup language,) and any other in fig 1 has been converted to *text* format. The total size of the collections used in this study was thirty-four point three megabytes (34).

- If my query is error the system displays the messages in fig 3: '**KanJechuubarbaaddankanadhaa**'? Message which is equivalent to "**Did You Mean**?" If the query which is s

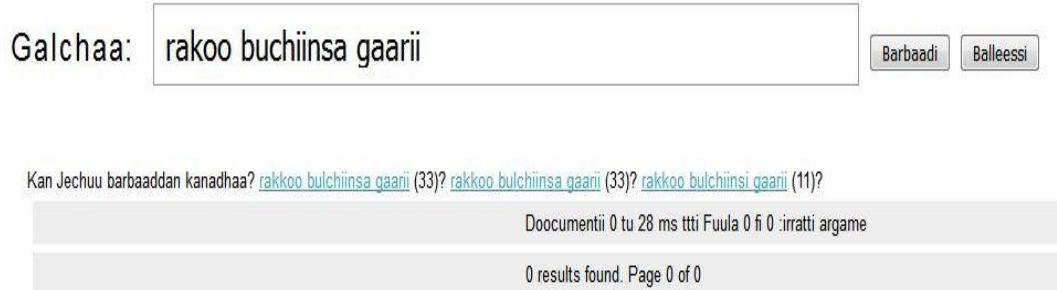

**Figure 2**: Response for did you mean

## Query selection

To show the performance of the system with the given corpus, query selection is very important. In this study 13 queries are randomly selected to check the performance of the system as shown in table 1. The average value of 13 queries is taken to judge the precision and recall of the system. The total number of documents we have been used were only 158, since it needs time and effort to collect more documents.

**Table 1**: query selection

| Num of queries | Queries | Number of documents which are: | |
|---|---|---|---|
| | | Relevant | Irrelevant |
| Query 1 | Rakkoo Bulchiinsa gaarii | 28 | 130 |
| Query 2 | Taphoota aadaa uummata oromoo | 3 | 155 |
| Query 3 | Ayyaana irreechaa | 5 | 153 |
| Query 4 | Macaafa qulqulluu | 5 | 153 |
| Query 5 | Aadaa uummata oromoo | 11 | 147 |
| Query 6 | Dorgommii guutuu Itoophiyaa | 7 | 151 |
| Query 7 | Sirna gadaa oromoo | 10 | 148 |
| Query 8 | Atileetota itoophiyaa | 7 | 151 |
| Query 9 | Geerarsa oromoo | 4 | 154 |
| Query 10 | Magaalaa guddoo Oromiyaa | 5 | 153 |
| Query 11 | Fayyaan faaya | 5 | 153 |
| Query 12 | Bilisummaa oromoo | 10 | 148 |
| Query 13 | Sochii diddaa gabrummaa | 7 | 151 |

## Method of System evaluation

It is known that measuring or evaluating the performance and accuracy of the system is very important after IR system is designed. According to [6], there are two main things to measure in IR system; these are: effectiveness of the system and its efficiency. In this study the effectiveness is measured by precision and recall. It is known that precision and recall are the main important ways of evaluating an IR system [7]. In addition to precision and recall, F-measure which is good for test of accuracy have been used.

$$Precision = \frac{number\ of\ retrived\ that\ are\ relevent}{total\ number\ of\ retrived\ documents}\ \ldots$$

$$Recall = \frac{number\ of\ retrived\ that\ are\ relevent}{total\ number\ relevent\ in\ the\ collection}\ \ldots$$

**Table 2:** Experimentation result before the lists of synonym and polysemy is added

| No Query | Queries | Total-Relevant In the collections | Total Retrieved | Relevant-Retrieved | Precision In % | Recall In % |
|---|---|---|---|---|---|---|
| Q 1 | Rakkoo Bulchiinsa gaarii | 28 | 33 | 28 | 84.84% | 100% |
| Q 2 | Taphoota aadaa uummata oromoo | 3 | 3 | 3 | 100% | 100% |
| Q 3 | Ayyaana irreechaa | 5 | 6 | 4 | 66.7% | 80% |
| Q 4 | Macaafa qulqulluu | 5 | 7 | 5 | 71.4% | 100% |
| Q 5 | Aadaa uummata oromoo | 11 | 16 | 11 | 68.75% | 100% |
| Q 6 | Dorgommii guutuu Itoophiyaa | 7 | 7 | 6 | 85.7% | 85.7% |
| Q 7 | Sirna gadaa oromoo | 10 | 17 | 8 | 47.05% | 80% |
| Q 8 | Atileetota itoophiyaa | 7 | 5 | 5 | 100% | 71.4% |
| Q 9 | Geerarsa oromoo | 4 | 8 | 4 | 50% | 100% |
| Q 10 | Magaalaa guddoo Oromiyaa | 5 | 7 | 4 | 57.14% | 80% |
| Q 11 | Fayyaan faaya | 5 | 5 | 4 | 80% | 80% |
| Q 12 | Bilisummaa oromoo | 10 | 15 | 8 | 53.3% | 80% |
| Q 13 | Sochii diddaa gabrummaa | 7 | 6 | 5 | 83% | 71.4% |
| Average | | | | | 72.91% | 86.8% |

Where,
*Total-Relevant* ➔ total number of documents that is relevant to the query in the collection
*Retrieved* ➔ total number of documents retrieved by the system
*Relevant-Retrieved* ➔ total number of retrieved documents that is relevant to the query

As it is shown in Table 2, the average recall and precision obtained from this study is 86.8% and 72.91% respectively. In the study, 100% (i.e. 1 when we change to integer) precision and recall is obtained for some queries. The reason for such event is explained as follows:

***Precision =1 (perfect precision)*** means every result returned by the search was relevant, but there may be other relevant documents that were not part of the research result. In other statement, perfect precision means every result retrieved by the search was relevant (but says nothing about whether all relevant documents were retrieved).

***Recall =1 (perfect recall)*** means all relevant documents were retrieved by the search but says nothing about how many irrelevant documents were also retrieved.

Recall measures how well the search system finds relevant documents; precision measures how well the system filters out the irrelevant documents, Precision is the ratio of the number of relevant documents retrieved to the total number retrieved. Recall is the ratio of the number of relevant documents retrieved to the total number of documents in the collection that are believed to be relevant. Recall considers the total number of relevant documents [8].

In addition to this, [7] also described that, as precision increases, recall decreases and vice versa. This is called a trade-off between precision and recall as they are inversely related which is also true as the result of this study shows. Depending on this, Query 5(*aadaa uummata oromoo*) registered better recall (100%) but lower precision (68.75%). This means that all relevant documents were retrieved by the system irrespective of the irrelevant document included in the result set to the given query. Additionally, there are many documents which talks about 'uummata Oromoo' (Oromo people) and all those documents are retrieved, but the relevant retrieved documents were not many. So, when small relevant retrieved documents (11) are divided by total retrieved documents (16), the result becomes low.

The result registered in Query 10 (*magaalaa guddoo oromiyaa*) in English (The capital city of Oromia) is also because of the word variation of the word 'guddoo' (capital). In Afaan Oromo the query '*magaalaa guddoo Oromiyaa*' can be written as: '*magaalaa beekamaa Oromiyaa*',' *magaalaa guddittii Oromiyaa*', '*magaalaa teessoo mootummaa naannoo Oromiyaa*' and so on. So, this can be improved when the lists of synonymy are inserted. Synonymy can increase recall by decreasing precision. Because, when lists of synonymy are added more number of documents are retrieved by the system.

### F -measure

F-measure was used so that the system increase recall by decreasing precision and sometimes vice-versa; there is a precision-recall tradeoff (for example, recall can be increased by simply retrieving more relevant documents, but the precision will go down). The F-measure combines precision and recall, taking their harmonic mean. The F-measure is high when both precision and recall are high. The formula is:

$$F - \text{measure} = \frac{2PR}{P + R}$$

Where; F-measure → stands for a measure of a test's accuracy

P -precision,

R- recall

The evaluation of the F-measure for the above average precision and recall is calculated as:

$$F = \frac{2(0.7291)(0.868)}{0.7291 + 0.868} = \frac{1.2657}{1.5971} = 0.7925$$

The F-measure assumes values in the interval [0,1]. It is 0 when no relevant documents have been retrieved, and 1 if all retrieved documents are relevant and all relevant documents have been retrieved. The F-measure result obtained for this study is 0.7925 which indicates that the accuracy of the system is 79.25%. Therefore, it is possible to have a search for Afaan Oromo text retrieval system using probabilistic approach.

## Effects of Synonyms and polysemy in Afaan Oromo

In most of natural languages including Afaan Oromo, the problem of synonym is very challenging task because, the absence of synonym has effect on the performance of the IR system. In this study we have made attempt to solve this problem for Afaan Oromo. There are two ways to specify synonym mappings in our collections. These are:

- ❖ Separate lists the two words with the symbol "=>" between them. If the token matches any word on the left, then the list on the right is substituted. The original token will not be included unless it is also in the list on the right.

**Example**: gaarii=>misha

- ❖ A comma-separated list of words. If the token matches any of the words, then all the words in the list are substituted, which will include the original token.

**Example**: gaarii, misha, bayeessa, dansa

Synonym mappings can be also used for spelling correction too.

**Example**: oromiya => Oromiyaa

Oromiya =>oromiyaa

ormiya => Oromiyaa

When there is no synonymy in the developed system, for example, when the query 'ormiya' is inserted, the system responds error message; but after we have inserted synonymy, the system displays all documents which contain the word 'oromiyaa'. So, it can improve the performance of the system.

Lists of Afaan Oromo synonymy (synonyms.txt) file are located under the folder C:\solr-6.0.0\server\solr\afaan-oromo-core\conf to add the synonym for our data. The following is where the lists of synonymy(synonymys.txt) were inserted.

```
<fieldType name="text_general" class="solr.TextField" positionIncrementGap="100">
<analyzer type="index">
<tokenizer class="solr.StandardTokenizerFactory"/>
```

```xml
<filter class="solr.StopFilterFactory" ignoreCase="true" words=" stopwords_ao.txt" />
<filter class="solr.LowerCaseFilterFactory"/>
</analyzer>
<analyzer type="query">
<tokenizer class="solr.StandardTokenizerFactory"/>
<filter class="solr.StopFilterFactory" ignoreCase="true" words=" stopwords_ao.txt" />
<filter class="solr.SynonymFilterFactory" synonyms="AOsynonyms.txt" ignoreCase="true" expand="true"/>
<filter class="solr.polysemyFilterFactory" synonyms="AOpolysemy.docx" ignoreCase="true" expand="true"/>

<filter class="solr.LowerCaseFilterFactory"/>
</analyzer>
</fieldType>
```

### Results after lists of synonymy are added to the document

The following table shows the effect of Afaan Oromo synonymy on system performance.

**Table 3:** Experimentation result after the addition of synonymy

| No. Query | Queries | Total-Relevant in the collections | Total Retrieved | Relevant-Retrieved | Precision in % | Recall in % |
|---|---|---|---|---|---|---|
| Q 1 | Rakkoo Bulchiinsa gaarii | 28 | 33 | 28 | 84.84% | 100% |
| Q 2 | Taphoota aadaa uummata oromoo | 3 | 3 | 3 | 100% | 100% |
| Q 3 | Ayyaana irreechaa | 5 | 6 | 4 | 66.7% | 80% |
| Q 4 | Macaafa qulqulluu | 5 | 7 | 5 | 71.43% | 100% |
| Q 5 | Aadaa uummata oromoo | 11 | 21 | 11 | 52.38% | 100% |
| Q 6 | Dorgommii guutuu Itoophiyaa | 7 | 7 | 6 | 85.7% | 85.7% |
| Q 7 | Sirna gadaa oromoo | 10 | 17 | 8 | 47.1% | 80% |
| Q 8 | Atileetota itoophiyaa | 7 | 5 | 5 | 100% | 71.4% |
| Q 9 | Geerarsa oromoo | 4 | 8 | 4 | 50% | 100% |
| Q 10 | Magaalaa guddoo Oromiyaa | 5 | 10 | 5 | 50% | 100% |
| Q 11 | Fayyaan faaya | 5 | 5 | 4 | 80% | 80% |
| Q 12 | Bilisummaa oromoo | 10 | 21 | 8 | 40% | 80% |
| Q 13 | Sochii diddaa gabrummaa | 7 | 7 | 7 | 100% | 100% |
| Average | | | | | 71.39% | 90.5% |

As it is shown in *Table 3* the addition of synonymy has effects on information retrieval system performance. As the result indicates after the synonymy is added the average number of recall is increased from 86.8% to 90.5% (i.e. 3.7% difference) and the average precision is decreased from 72.91% to 71.39% (i.e. 1.52% difference). This is because, when the lists of synonymy are added to the collections many documents can be retrieved. When many documents are retrieved and the relevant retrieved documents remain unchanged, the precision decreased. According to the experimentation made, the F-measure value after the addition of lists of synonymy is 79.82%, which is increased by 0.57 from the first. That means after the addition of synonymy the result is 79.82% and before the addition the result was 79.25%, so, the difference is 0.57. This shows that if standard thesaurus system is available for Afaan Oromo, the system performance is improved.

### Answers of research questions

As the result of the experiment made, the answer for the research question written under research question title is as given as follows.

**Question 1**: What are the effects of synonymy on Afaan Oromo Text Retrieval system?

**Answer:** Sometime it's difficult to memorize a word; hence synonymy plays an important role in such conditions and if we train the system (an Information Retrieval System) to use the synonymy of the word when the exact word is not provided for the search system. In this study we have prepared lists of synonymy which we call it as **AOsynonymy.txt,** and save it somewhere. Then when we search for some words in Afaan Oromo (for example, 'aadaa' which is 'culture' in English language, but if the word 'aadaa' is not exist in the collection and 'duudhaa' may exist) the IR system must return some documents instead of displaying the zero result, since the word 'aadaa' is the synonymy of 'duudhaa' in Afaan Oromo. Therefore, when we want to develop Afaan Oromo Text retrieval system, it is better to have the synonymy lists as much as possible.

**Example**: gaarii, misha, bayeessa, dansa

## 7. Conclusion

In this study we have attempted to develop a prototype of Afaan Oromo text retrieval system using probabilistic approach. The two modules for this development are indexing and searching. The text pre-processing or text operations are done in indexing. This is because to index the documents, the first thing we have to do is pre-processing. The searching component has also pre-processing and similarity measurement. The similarity measurement used was BM25similarity which is used in probabilistic approach and also applicable in apache solr.

According to the experimentation made in this study, the system registered the average precision and recall of 72.91% and 86.8% respectively, before the synonymy and polysemy were added. However, after the addition of some synonymy and polysemy, the result is changed to 71.39% average precision and 90.5% average recall for 13 queries of 158 documents. There were 3.7% improvements on recall and the precision was decreased by 1.52%. The F-measure (F-score) result for the first experimentation (before synonymy is added) was 79.25% and after the addition of synonymy the result was 79.82%, which was improved by 0.57% from the first. Therefore, we can say that the existence of both synonymy and polysemy have effect on the performance of an information retrieval system for Afaan Oromo. The probability that a relevant document is not retrieved (also called False Negative or Type II error) which is given by: **1- Recall** [9] is used to know the error rate of the system. Therefore, to get the probability that the system is unable to retrieve relevant document, we have to subtract the recall obtained by the experiment from one (1). That means; 1-0.905 =0.095, where 0.905 the result of recall in this study. So, for this study: The probability that the developed system can retrieve relevant document is =90.5% and the probability that the developed system cannot retrieve relevant document is =9.5%.

## Acknowledgement

First of all, I would like to thank my creator, 'GOD' for giving me health and patience to complete this work. '*Yaa waaqaayyo, barabaraan galatnikee hin xinnaatin*'. Next to this, I want to express my great gratitude to my friends.

## About the Author

**Isayas Wakgari Kelbessa**, the author of this paper is from Information system department, College of Computing and Informatics at Wolkite University, Ethiopia. He is graduated from Haramaya University, Ethiopia in B.Sc of Information Systems and M.Sc from Adama Science and Technology University, Ethiopia with the same department. Now he is a Lecturer at Wolkite University.